\documentstyle[aps,multicol]{revtex}
\renewcommand{\narrowtext}{\begin{multicols}{2} \global\columnwidth20.5pc}
\renewcommand{\widetext}{\end{multicols} \global\columnwidth42.5pc}

\input{epsf.tex}
\def\inseps#1#2{\def\epsfsize##1##2{#2##1} \centerline{\epsfbox{#1}}}

\begin{document}
\draft

\title{Skyrmions in Quantum Hall Systems with Realistic Force-Laws} 
\author{N.R. Cooper}
\address{Institut Laue-Langevin, Avenue des Martyrs, B.P. 156, 38042
Grenoble,
France.}
\date{16 September 1996}

\maketitle

\begin{abstract}
We study the charged excitations of quantum Hall systems at integer
filling fractions $\nu=2n+1$, for a force-law that takes account of
the finite width of the electron gas.  For typical values of this
width, in the limit of vanishing Zeeman energy we find that the
low-energy excitations are ``skyrmions'' not only at $\nu=1$ but also
at higher filling fractions.  Our results lead to the prediction that,
in typical samples, abrupt transitions to charged excitations with
very large spins should be observable at filling fractions higher than
$\nu=1$ if the Zeeman energy is reduced sufficiently.
\end{abstract}

\pacs{PACS Number: 73.40.Hm \hskip5.6cm {\it To appear in Physical Review B}}

\narrowtext

There has recently been a great deal of theoretical and experimental
activity studying spin-related phenomena in two-dimensional electron
systems in the quantum Hall
regime\cite{sondhi,fertig,macdonald,barrett,schmeller,aifer}.  In such
systems, the combination of strong Landau quantization and extremely
small Zeeman energy leaves the primary role in determining the
groundstate spin-correlations to the many-body interactions.
Attention has been focused predominantly on the filling fraction
$\nu=1$, at which interactions have particularly dramatic effects on
the spin-order.  Whilst the groundstate at this filling fraction is a
simple ferromagnet\cite{sondhi,macdonald} in which the electrons fill
a single, spin-polarized Landau band, the low-energy charged
excitations can have very unusual spin structures\cite{sondhi}.

At $\nu=1$, the picture that has emerged\cite{sondhi} is that, in the
strong-field limit (when the cyclotron energy is large compared to all
other energy scales), the form of the low-energy charged excitations
depends strongly on the ratio of the Zeeman energy $Z=g\mu_BB$ to the
exchange energy: this is proportional to
$e^2/4\pi\epsilon\epsilon_0\ell$ for Coulomb interactions
($\ell\equiv\sqrt{\hbar/eB}$ is the magnetic length in a magnetic
field $B$).  For large Zeeman energies, the charged excitations
involve the minimum number of spin-reversals: they are the ``polarized
quasiparticles'', with sizes of order the magnetic length and spins of
$1/2$.  In the opposite limit ($Z=0$), the charged excitations
minimize the exchange energy by adopting the ``skyrmion''
configurations\cite{sondhi} of the underlying continuum ferromagnet,
and have diverging spatial sizes and total spins.  At intermediate
values of the Zeeman energy, a compromise is reached in which the
charged excitations have finite sizes and spins, which are large
compared to the sizes and spins of the polarized
quasiparticles\cite{fertig}; following Ref.~\onlinecite{fertig} we
will refer to these large-spin particles as ``charged spin-textures''.
The low-energy charged excitations are predicted to be charged
spin-textures if the Zeeman energy is less than $0.054
(e^2/4\pi\epsilon\epsilon_0\ell)$\cite{rezayi,sondhi}, which is
typically the case in GaAs devices\cite{gaasfootnote}.  Experimental
studies of such systems show that the charged excitations at $\nu=1$
do carry large spin\cite{barrett,schmeller,aifer}, in good agreement
with current theory.

At higher odd filling fractions\cite{jain,wu}, the low-energy
properties in the strong-field limit depend only on the electrons in
the uppermost Landau band.  Consequently, such systems appear much
like a system at $\nu=1$, differing only in the form of the
single-particle states available to the electrons.  As is the case at
$\nu=1$, repulsive interactions lead to a ferromagnetic groundstate
for the uppermost band\cite{jain}.  However, for vanishing Zeeman
energy and for pure Coulomb interactions, the skyrmion excitations are
found to be {\it higher} in energy than the polarized quasiparticles
at filling fractions $\nu=3,5$ and $7$\cite{jain,wu}. It is therefore
believed that the low-energy charged excitations of quantum Hall
systems at filling fractions higher than $\nu=1$ are polarized
quasiparticles, even for vanishing Zeeman energy. The results of an
experiment probing the spins of the charged excitations at $\nu=3$ and
$5$ are indeed consistent with spin-polarized
excitations\cite{schmeller}.
 
Although the current experimental observations at $\nu=2n+1$ are in
qualitative agreement with existing theory, there do remain important
unanswered questions concerning the properties of the charged
excitations.  In particular, there has been no systematic study of the
effects of the finite thickness of the electron layer on these
particles (the one study\cite{xie} to have taken this into account
considered only $\nu=1$).  The purpose of the present work is to
investigate how the above theoretical picture, established assuming
{\it pure} Coulomb interactions, is affected when a more realistic
force-law is used.  We find that {\it qualitative} differences arise:
in the limit of vanishing Zeeman energy, skyrmions are found to be the
low-energy excitations not only at $\nu=1$ but also at higher filling
fractions.  We will concentrate on the results of our calculations;
details will be presented elsewhere\cite{cooper}.

For the most of this work, we study the properties of the charged
excitations at integer filling fractions $\nu=2n+1$ in the
strong-field limit, for the force-law (from now on we express energies
in units of $e^2/4\pi\epsilon\epsilon_0\ell$ and lengths in units of
the magnetic length $\ell$)
\begin{equation}  
V_w(r) \equiv
\int\!\!\int_{-\infty}^\infty \!\!\!\!\!  dz dz^\prime
\frac{e^{-\frac{z^2+{z^\prime}^2}{2w^2}}}{2\pi w^2}
\frac{1}{\sqrt{r^2+(z-z^\prime)^2}}.
\label{eq:forcelaw}
\end{equation}  
This represents the interaction between electrons with Gaussian
subband wavefunctions of width $w$.  Although subband wavefunctions
are not precisely of this form, the force-law (\ref{eq:forcelaw})
captures the essential effect of the finite width of the electron
layer: the softening of the repulsive interactions at short distances.
Comparison with experimental systems may be achieved by equating $w$
to the r.m.s.  width of the subband charge density, which is typically
of the order of the magnetic length\cite{zhang}.

The main consequences of the finite width $w$ at a filling fraction
$\nu=2n+1$ may be understood by comparing the energy gap for the
creation of a widely-separated quasielectron/quasihole pair to that of
a skyrmion/antiskyrmion pair, in the limit of vanishing Zeeman
energy\cite{polarizedfootnote}.  The quasiparticle gap is set by the
exchange energy\cite{kallin}, which may be written
\begin{equation}  
E_{qe/qh} = \frac{1}{2\pi}\int_0^\infty \!\!\!\!  dq\; q {\tilde
V}_w(q)\left[L_n(q^2/2)\right]^2 e^{-q^2/2},
\label{eq:qpgap}
\end{equation}  
where ${\tilde V}_w(q)=(2\pi/q)e^{q^2w^2}\mbox{erfc}(qw)$ is the
Fourier transform of the force-law (\ref{eq:forcelaw}), and $L_n(z)$
is a Laguerre polynomial with degree $n$ equal to the index of the
uppermost Landau band.  The skyrmion gap is set by the spin-wave
stiffness\cite{kallin}, and may be expressed as
\begin{equation}
E_{sk/ask} =\frac{1}{2\pi}\int_0^\infty \!\!\!\!  dq \; (q^3/2)
{\tilde V}_w(q)\left[L_n(q^2/2)\right]^2 e^{-q^2/2}.
\label{eq:skgap}
\end{equation}

\begin{figure}
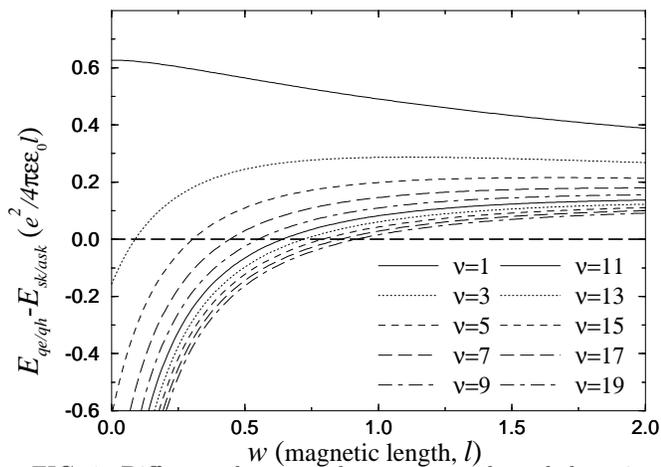

\inseps{fig1.eps}{0.5}
\caption{Difference between the quasiparticle and skyrmion energy gaps
for vanishing Zeeman energy at various filling fractions $\nu$, as a
function of the layer thickness $w$.  A positive energy difference
indicates that the skyrmion/antiskyrmion pair has a lower energy than
the quasielectron/quasihole pair.}
\label{fig:gaps}
\end{figure}

Figure~\ref{fig:gaps} shows the difference between these two gaps as a
function of the layer thickness $w$.

For Coulomb interactions ($w=0$), the skyrmion gap is less than the
quasiparticle gap only at $\nu=1$.  It is just this comparison that
was made by Wu and Sondhi\cite{wu}, and that lead them to conclude
that, while skyrmions are the low-energy charged excitations at
$\nu=1$ for vanishing Zeeman energy, this is {\it not} the case at
higher odd filling fractions.

However, as can be seen from Fig.~\ref{fig:gaps}, only a very small
width ($w\gtrsim 0.09\ell$) is required before the skyrmions become
lower in energy than the polarized quasiparticles at $\nu=3$. For a
sample with $w=\ell$, skyrmions are the lower-energy excitations at
all filling fractions up to and including $\nu=21$.  The sensitivity
of $E_{qe/qh}-E_{sk/ask}$ to changes in $w$ is primarily due to the
strong dependence of the skyrmion energy (\ref{eq:skgap}) on the
short-range force-law.  That even a small value of $w$ is sufficient
to cause the skyrmions to be the lower energy excitations is one of
the principal observations of the present work.

In view of the fact that small changes in the force-law can have
significant qualitative consequences, one might expect that the
screening arising from Landau level mixing (which we have excluded
until now by considering the strong-field limit) could also be
important.  We have studied the effects of Landau level mixing on the
skyrmion (\ref{eq:skgap}) and quasiparticle (\ref{eq:qpgap}) gaps at
$\nu=2n+1$, within an RPA approach in which the interaction
(\ref{eq:forcelaw}) is screened by a static wavevector-dependent
dielectric constant arising from the particle/hole excitations of the
$2n+1$ filled Landau bands\cite{algla}.  At filling fractions higher
than $\nu=1$ we find that the values of the minimum subband widths
above which skyrmions are lower in energy than polarized
quasiparticles are increased by such screening. Thus, Landau level
mixing appears to favour the polarized quasiparticles over the
skyrmions.  Nevertheless, for the situation in which the cyclotron
energy is equal to $e^2/4\pi\epsilon\epsilon_0\ell$ ($B=6.8\mbox{T}$
in GaAs\cite{gaasfootnote}), we find that when $w=\ell$ the skyrmions
are the lower-energy excitations at all filling fractions up to and
including $\nu=7$.  While a more detailed study of Landau level mixing
is clearly desirable, the RPA results suggest that our qualitative
conclusion is not affected: for a typical sample, skyrmions would have
lower energies than polarized quasiparticles at filling fractions
higher than $\nu=1$ if the Zeeman energy were sufficiently small.

We now turn to discuss the properties of the charged excitations at
non-zero Zeeman energy.  We return to considering the strong-field
limit in which Landau level mixing is negligible. The qualitative
behaviour that we find is not affected by the inclusion of a small
amount of Landau level mixing in the manner described above.

We have studied the charged excitations at finite Zeeman energy within
the Hartree-Fock (HF) approximation introduced by Fertig {\it et
al.}\cite{fertig}, generalizing this approach to higher filling
fractions $\nu=2n+1$, and to the force-law (\ref{eq:forcelaw}).  To
restrict to a finite number of single-particle basis states (which is
necessary for the numerical solution of the HF approach), we consider
the electrons to lie on the surface of a sphere pierced by $2S$ flux
quanta, for which the single-particle states in the Landau level with
index $n$ can have $\hat{\bbox z}$-component of the angular momentum
in the range $m=-S-n, -S-n+1,\ldots ,S+n$\cite{sphere}.  The HF
wavefunction for the positively-charged excitation may then be written
\begin{equation}  
|HF\rangle \equiv
\prod_{m=-S-n}^{S+n-1}\!\!\!\!  (u_m c_{nm\downarrow}^\dagger +v_m
c_{nm+1\uparrow}^\dagger)|0\rangle,
\label{eq:hfwavefunction}
\end{equation}  
where $|0\rangle$ is the vacuum state, and $c_{nm\sigma}^\dagger$
creates an electron with spin $\sigma$ in the state with Landau level
index $n$ and angular momentum $m$.  The HF procedure involves the
minimization of the average energy with respect to the variational
parameters $\{u_m,v_m\}$, subject to the constraint that the
wavefunction is normalized. This leads to a set of self-consistent
equations\cite{cooper}, of the same form as those presented in
Ref.~\onlinecite{fertig}, which may be iterated until convergence is
achieved.  The results that we will present are based on calculations
with system-sizes up to $2S=200$.

\begin{figure}
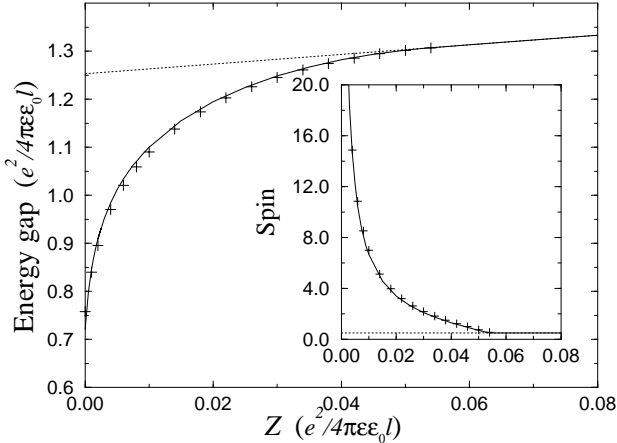

\inseps{fig2.eps}{0.48}
\caption{Energy gap as a function of the Zeeman energy at $\nu=1$ and
$w=0$, from Hartree-Fock calculations on a sphere with $2S=200$ (solid
line), and $2S=100$ (crosses).  The dotted line is the result for
polarized quasiparticles.  The spin of one charged excitation is shown
in the inset.}
\label{fig:n0w0}
\end{figure}

We will first discuss our results for the filling fraction $\nu=1$.
Results for the case of Coulomb interactions ($w=0$) are presented in
Fig.~\ref{fig:n0w0}.  This figure shows the energy-gap to the creation
of a widely-separated pair of particles with opposite charges, and the
spin of one of these particles (due to an exact particle/hole
symmetry, both particles have the same spin).  We define the spin of
an excitation to be the change in spin when it is introduced to the
spin-polarized filled Landau band: the polarized quasiparticle
therefore has spin 1/2.

For small Zeeman energies, the energy gap is less than that for
polarized quasiparticles; in this regime, the lowest-energy HF state
is a charged spin-texture with a large spin.  At large Zeeman
energies, the lowest-energy HF state is the polarized quasiparticle
($|u_m|^2=1-|v_m|^2=0$), so the energy gap and spin become equal to
those for the quasiparticles.  Defining a critical Zeeman energy
$Z^{HF}_c$ by the value of $Z$ at which the transition between the
charged spin-texture and the polarized quasiparticle occurs, we find
$Z^{HF}_c=0.054$.

Our qualitative results are very similar to those presented in
Ref.~\onlinecite{fertig}, where the same situation ($\nu=1$, $w=0$)
was studied in the disc geometry.  However, there are quantitative
differences: the energy gaps and spins that we find are consistently
less than those of Ref.~\onlinecite{fertig}.  These discrepancies may
be due to larger finite-size effects on the disc than on the sphere.
Finite-size effects are small in our calculations, as may be seen by
comparing the results for $2S=100$ and $200$ in Fig.~\ref{fig:n0w0}.

We now introduce a non-zero width $w$. We find no changes in the
qualitative behaviour shown in Fig.~\ref{fig:n0w0}. There are, however
quantitative corrections: at each (non-zero) Zeeman energy, the energy
and spin of the charged spin-texture are closer to those of the
polarized quasiparticle.  For example, at $Z=0.01$ the spin decreases
from 6.7 for $w=0$ to 5.9 for $w=0.5\ell$, and 5.0 for $w=\ell$; at
$Z=0.02$ there is a decrease from 3.4 ($w=0$) to 3.0 ($w=0.5\ell$),
and 2.4 ($w=\ell$).  The critical Zeeman energies are also reduced
[see Table~\ref{table}].  Similar qualitative behaviour has been
observed in exact-diagonalization studies of small systems\cite{xie}.

\begin{figure}
\inseps{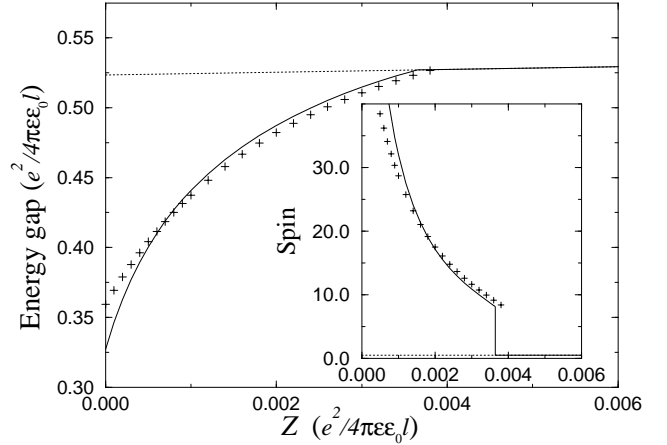}{0.48}
\caption{Same as Fig.~\protect\ref{fig:n0w0} for $\nu=3$ and $w=\ell$.}
\label{fig:n1w1}
\end{figure}

Much more interesting behaviour occurs for the higher filling
fractions.  Fig.~\ref{fig:n1w1} shows our results at $\nu=3$ and
$w=\ell$.  Although we do not present detailed results here, the same
qualitative behaviour is found for all higher filling fractions and
values of $w$ that we have studied ($3\leq\nu\leq 7$, $w\leq 3$),
provided that for $Z=0$, $w$ is sufficiently large that the skyrmion
gap (\ref{eq:skgap}) is less than the quasiparticle gap
(\ref{eq:qpgap}).  Finite-size effects are larger in
Fig.~\ref{fig:n1w1} than in Fig.~\ref{fig:n0w0}, due to the much
larger spins and spatial extents of the charged spin-textures.

As is the case for $\nu=1$, with increasing Zeeman energy the spin of
the charged spin-texture decreases and there is a critical value
$Z^{HF}_c$ above which the lowest-energy state is the polarized
quasiparticle.  However, in contrast to the continuous behaviour found
for $\nu=1$, for the higher filling fractions we observe a {\it
discontinuous} transition at $Z^{HF}_c$ from a charged spin-texture
with a large spin to the polarized quasiparticle.  The average spin of
the charged spin-texture at the transition $S^{HF}_c$ is given in
Table~\ref{table}.

We note that the distinction between the continuous transition at
$\nu=1$ (in which the spin falls smoothly to 1/2 at $Z=Z^{HF}_c$) and
the first-order transition at higher filling fractions (in which the
spin jumps from $S^{HF}_c\neq 1/2$ to 1/2 at $Z=Z^{HF}_c$) is an
artifact of the HF approximation.  In an exact treatment, these
transitions must {\it both} be discontinuous, since both result from
level crossings in which there are changes in the spin quantum number.
[The wavefunction (\ref{eq:hfwavefunction}) is not an eigenstate of
spin, so the {\it average} spin of the HF groundstate can vary
continuously.]  Nevertheless, the HF results do indicate that the
nature of the level crossings differ qualitatively at $\nu=1$ and at
higher filling fractions.

At $\nu=1$, the continuous reduction in spin suggests that the
transition to the polarized quasiparticle occurs from the smallest
possible charged spin-texture, which has spin-3/2.  This is confirmed
by the good agreement [see Table~\ref{table}] between $Z_c^{HF}$ and
the value $Z^{(3/2)}_c$ at which the spin-3/2 charged spin-texture has
the same energy as the polarized quasiparticle (we obtain the values
$Z^{(3/2)}_c$ from exact-diagonalization studies on the
sphere\cite{cooper}).  The slight deviations can be accounted for by
the fact that the HF approach is not exact.

At higher filling fractions, the HF results indicate that the
transition occurs from a charged spin-texture with a spin much larger
than 3/2. Indeed, we find that $Z^{HF}_c$ is consistently {\it larger
than} $Z^{(3/2)}_c$.  Thus, when $Z=Z^{(3/2)}_c$ (such that the
polarized quasiparticle and the spin-3/2 charged spin-texture have
equal energies), the HF results show that there exists a lower-energy
charged spin-texture, with a spin larger than $3/2$.  This confirms
that the transition involves a change of spin that is larger than
unity.

We now turn to discuss the experimental implications of our results.

For $\nu=1$, the reduction of finite-size effects and the introduction
of the thickness $w$ change the quantitative results of Fertig {\it et
al.}\cite{fertig}. At any given (non-zero) value of the Zeeman energy,
the spin of the charged spin-texture is reduced.  This brings the HF
results into better agreement with experimental
measurements\cite{barrett,schmeller,aifer}.

For higher filling fractions, we find that, in typical samples,
charged spin-textures would be lower in energy than the polarized
quasiparticles if the Zeeman energy were sufficiently small.  However,
the critical values below which the Zeeman energy must lie in order
that these novel particles appear are very small: they are about
$1/10$ of the critical Zeeman energies at $\nu=1$ [see
Table~\ref{table}].  This condition is not typically achieved in GaAs
devices. In particular, the range of Zeeman energy studied in the
experiments of Schmeller {\it et al.}\cite{schmeller} ($Z\gtrsim
0.0075$) lies above the critical Zeeman energies that we predict for
their sample ($w\simeq 0.5\ell$) at $\nu=3$ and $5$: the observation
of spin-polarized charged excitations under these
conditions\cite{schmeller} is therefore consistent with our results.

It may be difficult to bring the Zeeman energy below the critical
values we predict by simply reducing the magnetic field and carrier
density, owing to the increasing effects of disorder. To observe
charged spin-textures at filling fractions higher than $\nu=1$, it may
be necessary to reduce the Zeeman energy by other means.  This can
best be done through the application of external pressure, which can
be used to tune the Zeeman energy to zero\cite{nicholas}.  Our results
lead to the prediction that, as the Zeeman energy of a typical sample
is reduced from its value at ambient pressure, the low-energy charged
excitations at filling fractions higher than $\nu=1$ will undergo
abrupt transitions from polarized quasiparticles to charged
spin-textures with very large spins.

\vskip0.2cm
The author is grateful to B.I. Shklovskii, J.J. Palacios and D.E.
Logan for helpful discussions and comments, and to the Aspen Center
for Physics for hospitality.

\newpage

\begin{table}
\begin{tabular}{c||c|c||c|c||c|c}
 & \multicolumn{2}{c||}{$\nu=1$} & \multicolumn{2}{c||}{$\nu=3$} &
\multicolumn{2}{c}{$\nu=5$} \\ \hline 
$w$ & $Z^{HF}_c$ [$S^{HF}_c$] &  $Z^{(3/2)}_c$ & $Z^{HF}_c$[$S^{HF}_c$] &  $Z^{(3/2)}_c$ 
& $Z^{HF}_c$[$S^{HF}_c$] &  $Z^{(3/2)}_c$ \\ \hline
0.0 & .054 [.5] & .054  & N/T & N/T & N/T & N/T\\
0.5 & .048 [.5] & .049 & .0018 [20.1] & N/T & .00012 [83] & N/T \\
1.0 & .038 [.5]& .040 & .0037  [8.1] & .0029 & .0011 [25.0] & .00017 \\
2.0 & .026 [.5]& .027  & .0044  [4.1] & .0043 & .0018
 [11.0] & .0014 \\
\end{tabular}
\caption{Zeeman energies $Z^{HF}_c$ (in units of 
$e^2/4\pi\epsilon\epsilon_0\ell$) at which the Hartree-Fock solution
undergoes a transition from a charged spin-texture (with average spin
$S^{HF}_c$) to a polarized quasiparticle.  $Z^{(3/2)}_c$ is the Zeeman
energy at which the polarized quasiparticle has the same energy as the
spin-3/2 charged spin-texture.  N/T indicates that there is no
transition.}
\label{table}
\end{table}

\vskip20cm

\widetext


\begin{thebibliography}{10}
\bibitem{polarizedfootnote}
We assume the groundstate of the uppermost Landau band to be
ferromagnetic for all $\nu$ and $w$ that we consider.
\end{thebibliography}
\end{document}